\documentclass
[aps,prl,twocolumn,superscriptaddress,showpacs,floatfix]{revtex4-1}%
\usepackage{graphicx,amsmath,amssymb}
\usepackage{amsmath}
\usepackage{amsfonts}
\usepackage{amssymb}
\usepackage{graphicx}%
\providecommand{\U}[1]{\protect\rule{.1in}{.1in}}
\begin{document}
\title{Real-time and -space visualization of excitations \\of the $\nu = 1/3$ fractional quantum Hall edge}
\author{Akinori Kamiyama}
\affiliation{Department of Physics, Tohoku University, Sendai 980-8578, Japan}
\author{Masahiro Matsuura}
\affiliation{Department of Physics, Tohoku University, Sendai 980-8578, Japan}
\author{John N. Moore}
\affiliation{Department of Physics, Tohoku University, Sendai 980-8578, Japan}
\author{Takaaki Mano}
\affiliation{National Institute for Materials Science, Tsukuba, Ibaraki 305-0047, Japan}
\author{Naokazu Shibata}
\affiliation{Department of Physics, Tohoku University, Sendai 980-8578, Japan}
\author{Go Yusa}
\email{yusa@tohoku.ac.jp}
\affiliation{Department of Physics, Tohoku University, Sendai 980-8578, Japan}
\affiliation{ 
Center for Spintronics Research Network, Tohoku University, Sendai 980-8577, Japan}
\date{\today
}

\begin{abstract}
     We present scanning optical stroboscopic confocal microscopy and spectroscopy measurements wherein three degrees of freedom, namely energy, real-space, and real-time, are resolvable. The edge-state propagation is detected as a temporal change in the optical response in the downstream edge. We succeeded in visualizing the excited states of the most fundamental fractional quantum Hall (FQH) state and the collective excitations near the edge. The results verify the current understanding of the edge excitation and also point toward further dynamics outside the edge channel.

\end{abstract}

\maketitle

When electrons confined to two dimensions (2D) are subjected to a perpendicular magnetic field $B$, and the ratio of their density $n_e$ and the flux quanta density (i.e. $\nu$, the Landau-level filling factor) is an integer or rational fraction, the bulk of the system develops a gap to form integer and fractional quantum Hall (QH) states. Herein, $\nu = \frac{h}{eB} n_e$, where $h$ and $e$ denote the Planck's constant and elementary charge, respectively. The QH system is the most extensively studied 2D topological material\cite{thouless, tsui, wen, hasan} wherein the bulk of the system comprises an insulator, and electric conduction is restricted to the edge. Because $B$ breaks the time-reversal symmetry, the one-dimensional (1D) edge current, i.e. edge-state excitations called edge magneto-plasmons (EMPs)\cite{aleiner}, propagates unidirectionally. 
In contrast to the bulk, the system edge contains compressible regions with a gapless excitation. This QH edge state constitutes a 1D system that exhibits uniquely high coherence\cite{ji} 
and hosts a wide variety of physics\cite{bid, sabo, venkatachalam, ashoori, ernst, kamata, hashisaka}, such as the Tomonaga-Luttinger liquid\cite{kamata, hashisaka}, anyonic statistics
\cite{nakamura, bartolomei}, and charge-neutral upstream Majorana modes of the edge current\cite{bid}. 
Furthermore, it offers the potential for exotic applications, including topological and flying qubit quantum computation
\cite{kitaev, sarma, shimizu} and quantum energy teleportation
\cite{matsuura, yusaPRA, hotta}. 
The properties of these edge states are typically examined through electric measurements
\cite{ji, bid, sabo, venkatachalam, ashoori, ernst, kamata, hashisaka, shimizu, matsuura, feve}, including shot-noise
\cite{ji, bid, sabo} and heat-transport measurements\cite{venkatachalam}. 
The edge current dynamics have been specifically examined by time-resolved transport measurements
\cite{ashoori, ernst, kamata, hashisaka, matsuura}. 
In general, measurements that detect electric signals from electrodes attached to a sample are limited in that they only obtain information from a certain fixed region. 
By contrast, scanning optical microscopy can quantify the properties of the electronic system, such as spin polarization, in the spatial domain
\cite{hayakawa, moore}. 
This is because the intensity and photon energy of photoluminescence (PL) emitted from the bound states of two electrons and a hole, known as charged excitons or trions, are highly sensitive to the electronic environment around the trions
\cite{hayakawa, moore, yusaPRL} [see Supplementary Material (SM)]. 

\begin{figure}
    %orginal EPS size 0 0 241 171
    %\includegraphics[bb=18 32 225 152, clip,width=8.6cm]{Fig1.eps}
    %\includegraphics[bb=18 30 208 138, clip,width=8.6cm]{Fig1.eps}
    %\includegraphics[bb=0 0 219 124, clip,width=8.6cm]{Fig1.eps} \caption{}%
    %\includegraphics[bb=0 0 219 123, clip,width=8.6cm]{Fig1.eps}
    %\includegraphics[clip,width=8.6cm]{Fig1.eps}
    \par
    \begin{center}
    \includegraphics[clip]{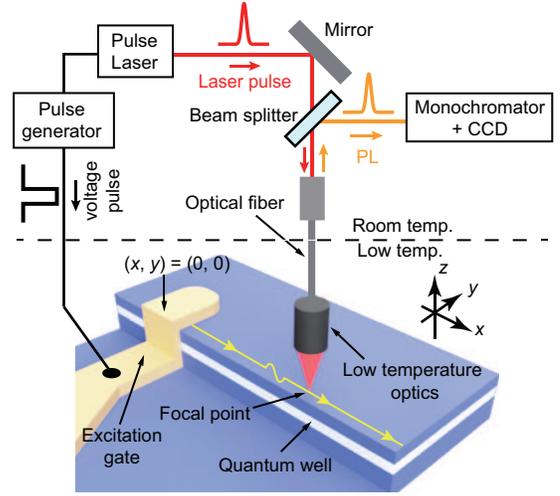}
    \end{center}
\caption{(Color online) Schematic of confocal scanning stroboscopic microscopy and spectroscopy (see SM for details). 
The origin in the rectangular coordinates is defined as $x=0$ at the center of the excitation gate and $y=0$ at the sample edge. The direction of chirality determined by the direction of Lorentz force acting on the electrons by $B$ is indicated by the yellow arrows. All measurements presented were performed at $B=14$ T at $\sim 40$ mK unless otherwise specified. }%
\label{fig:fig1}%
\end{figure}

\begin{figure*}[t]
    \par
    \begin{center}
    \includegraphics[clip]{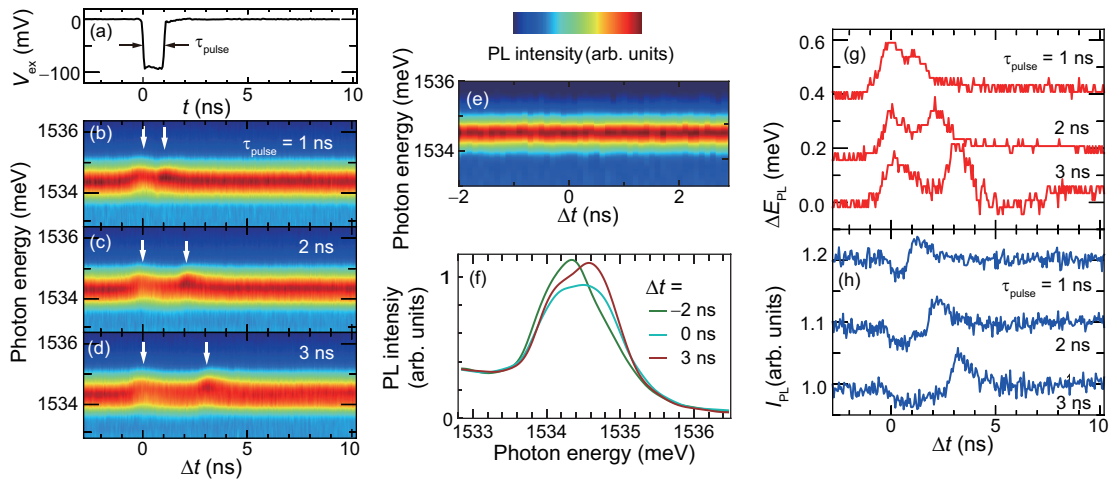}
    \end{center}
    \caption{(Color online) (a) 
    Waveform of $V_{\mathrm{ex}}$ for the pulse duration $\tau _{\mathrm{pulse}}=1$~ns measured using an oscilloscope with a $6$~GHz bandwidth by direct connection to the pulse generator. 
    The typical rise and fall times are $\sim 60$~ps. The actual voltage form applied to the excitation gate is different from that of $V_{\mathrm{ex}}$, because of the insertion losses in the dilution refrigerator. 
    (b), (c), and (d) PL spectra downstream at point 1 $(x,y)=(17$~$\mu$m$,1$~$\mu$m$)$ as a function of $\Delta t$ at $\tau _{\mathrm{pulse}}=1$, $2$, and $3$~ns, respectively (see SM for the sample geometry).
    (e) PL spectra upstream at point 2 $(-18$~$\mu$m$,1$~$\mu$m$)$ as a function of $\Delta t$ for $\tau _{\mathrm{pulse}}=1$~ns obtained at $\sim 100$~mK.
    (f) PL spectra obtained at $\tau _{\mathrm{pulse}}=3$~ns and $\Delta t=-2$, $0$, $3$~ns. 
    (g) Energy shift $\Delta E_{\mathrm{PL}}$ of the PL peak $E_{\mathrm{PL}}$ as a function of $\Delta t$ obtained from the data in Figs. 2(b), 2(c), and 2(d). The $y$-axis is offset for clarity.
    (h) PL intensity, $I_\mathrm{PL}$, integrated from $1532.9$ to $1536.5$~meV as a function of $\Delta t$ calculated from the data in Figs. 2(b), 2(c), and 2(d). The $y$-axis is offset for clarity.
}%
    \label{fig:fig2}%
    \end{figure*}

To explore the edge-state dynamics in real space and time, we performed scanning confocal microscopy and spectroscopy experiments using a strobe effect technique known as temporal aliasing (Fig. 1) Our scanning confocal microscopy apparatus collects light from illuminated 2D electrons in a 15-nm GaAs/AlGaAs quantum well (QW) (see SM details regarding the sample).
A mode-locked Ti:sapphire laser pulse is transmitted through an optical fiber and low-temperature optics, and focused on a selected position of the sample to microscopically excite trions. 
The resulting microscopic PL spectra are collected through the objective lens and fiber, and then sent to a spectrometer and measured using a charge-coupled device (CCD) detector. 
The low-temperature optics partially depicted in Fig. 1 comprise mirrors, a polarizing beam splitter, a $\lambda /4$ wave plate, and an objective lens. They are located near the sample, and are used to focus the laser and selectively collect the $\sigma ^-$-polarized PL.
The PL collected from the sample is transmitted through the same fiber and introduced to the monochromator and CCD. The focal point of the confocal microscope can be aligned to the sample surface and scanned across the sample in 2D by piezoelectric stages (See SM for details for the scanning optical microscopy).

\begin{figure*}[t]
    \par
    \begin{center}
    \includegraphics[clip]{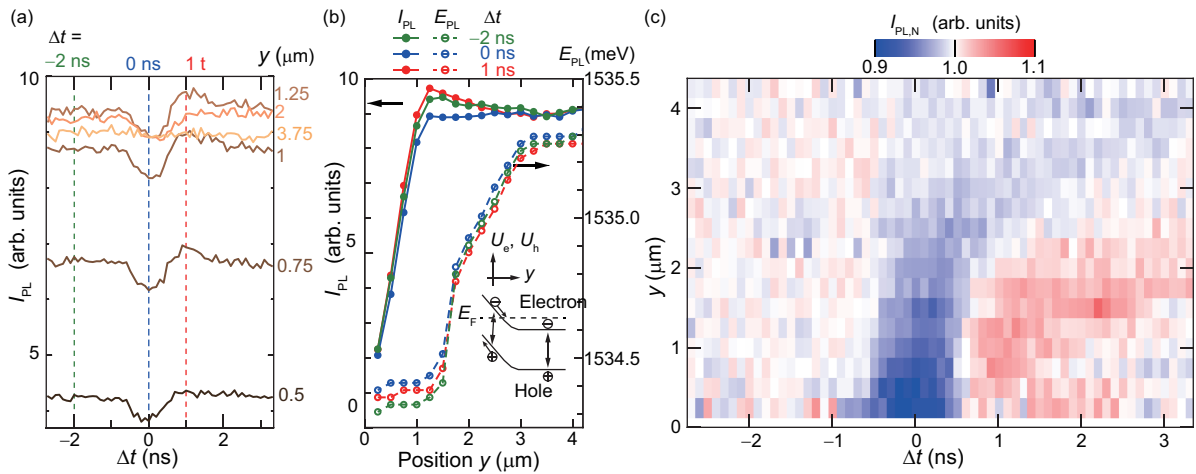}
    \end{center}
    \caption{(Color online) (a) $\Delta t$ dependence of $I_{\mathrm{PL}}$ ingetated between $1533.1$ and $1536.5$~meV at different values of $y$. $\tau _{\mathrm{pulse}}$ is $1$~ns. (b) Spatial dependence of $I_{\mathrm{PL}}$ and $E_{\mathrm{PL}}$ at three different values of $\Delta t$. 
    (Inset) Schematic of the band structure. The confinement potential $U_{\mathrm{e}}$ and $U_{\mathrm{h}}$ for the electrons and holes, respectively, are shown as a function of $y$. 
    (c) Spatial and temporal dependence of normalized integrated PL intensity $I_{\mathrm{PL,N}} (\Delta t, y)$ in $y$ and $\Delta t$. (see body text for definition.)}%
    \end{figure*}
    
A laser pulse is synchronized to a voltage pulse applied to the excitation gate deposited on top of the sample\cite{matsuura} (Figs. 1, 4(a), and S1) with a controllable time delay $\Delta t$. 
This ensures that the measured PL reflects the electronic state at a moment following the voltage pulse. This pulse creates a local charge disturbance that propagates as an EMP wave packet
\cite{ashoori, ernst, kamata, hashisaka, matsuura} along the boundary of the QH system defined by the interface of the QW and the vacuum. For a specific $\Delta t$, the wave packet arrives at the laser focal point in coincidence with the laser pulse illumination. 
This allows us to capture the PL spectra reflecting the electronic state modified by the wave packet. We measure the PL as a function of $\Delta t$ and the spatial coordinates of the laser spot parallel (the $x$-axis) and perpendicular (the $y$-axis) to the edge.

In this study, we focus on the $\nu = 1/3$ fractional QH (FQH) state, which is one of the most fundamental Laughlin states\cite{tsui}, realized at a temperature of $\sim 40$~mK, where the longitudinal resistance of our sample $R_{xx}$ reaches approximately zero [See SM and Fig. S2(b)].

First, we present the time delay dependence of the PL spectra observed at fixed spatial points. 
The wave packet is excited by a square pulse of the duration $\tau _{\mathrm{pulse}}$ [Fig. 2(a)]. 
When $\tau _{\mathrm{pulse}}=1$~ns, the PL peak energy is blue-shifted at two time positions $\Delta t$ of $\sim 0$~ns and $\sim 1$ ns $17$~$\mu$m (point 1 in Fig. S1 in SM) downstream from the excitation gate [Fig. 2(b)].

Herein, $\Delta t=0$ is defined by the moment of the first blue-shift. As $\tau _{\mathrm{pulse}}$ is extended to $2$ and $3$~ns, the second blue-shift becomes delayed by the same amount [Figs. 2(c) and 2(d)], thus indicating that these shifts correspond to the fall and rise of the square pulse. 
Notably, no significant changes are observed in the PL spectra [Fig. 2(e)] $18$~$\mu$m upstream from the excitation gate (point 2 in Fig. S1). Therefore, the stroboscopically observed changes in PL are induced by the electrically excited wave packet as it chiraly propagates along the edge. 
The PL spectrum [Fig. 2(f)] observed at $\Delta t=-2$~ns, observed before the first blue-shift, contains one intense peak; whereas the spectra during the blue shifts observed at $\Delta t=0$ and $3$ ns appear to contain two peaks.
The PL spectra at $\nu \sim 1/3$ may contain at most three peaks originating from the singlet and two triplet trions\cite{yusaPRL}. These triplets, which can be classified as dark and bright triplets, are distinguished by their total angular momentum\cite{wojs}, and exhibit different $B$ and $n_e$ dependences\cite{yusaPRL}. 
At $B \sim 14$~T (this work), the singlet PL intensity is too low to be detected (Fig. S2), whereas the two triplet peaks almost overlap. Therefore, the overlapping peaks observed with the blue-shifts at $\Delta t=0$ and $3$~ns [Fig. 2(f)] may be assigned to the dark and bright triplets, respectively, corresponding to the low- and high-energy peaks. 
Herein, we define the size of the PL peak energy shift as $\Delta E_{\mathrm{PL}}$ 
and the integrated PL intensity as 
$I_{\mathrm{PL}}$ [Figs. 2(g) and 2(h)]. 
The two peaks in the waveform of 
$\Delta E_{\mathrm{PL}}$ as a function of $\Delta t$ are $~200$~$\mu$eV; whereas $I_{\mathrm{PL}}$ displays a dip followed by a peak, coinciding with the peaks in 
$\Delta E_{\mathrm{PL}}$. Notably, neither the waveforms of $\Delta E_{\mathrm{PL}}$ nor $I_{\mathrm{PL}}$ resemble the waveform of $V_{\mathrm{ex}}$; instead, the $I_{\mathrm{PL}}$ waveform strongly resembles $(dV_{\mathrm{ex}})/dt$. We confirmed the resemblance by repeating the experiment using sawtooth voltage pulses with a fast rise or fall (See Fig. S3 in SM). Moreover, in a previous study, which also excited wave packets by a front gate in the $\nu =1$ integer QH edge, waveforms resembling $(dV_{\mathrm{ex}})/dt$ were detected by another front gate\cite{matsuura}.

The observation that the PL spectrum changes only in response to $(dV_{\mathrm{ex}})/dt$ is consistent with the process of wave packet generation by the metal gate. 
Consider the application of the square pulse in Fig. 2(a). 
Because the metal gate is capacitively coupled to the edge state, during the fall of $V_{\mathrm{ex}}$, a wave packet is created, propagating along the edge to the optical focal point, and is detected in the PL spectrum. Once $V_{\mathrm{ex}}$ reaches a minimum value, the edge current flows along the new potential profile created by the constant $V_{\mathrm{ex}}$. 
Thus, the edge state is not excited during the flat part of the pulse, and the PL spectrum appears as it did in the absence of the wave packet. During the rising part of $V_{\mathrm{ex}}$, a wave packet with opposite charges is created that propagates along the edge, again resulting in a change in the PL spectrum at the focal point. 
The waveform of $I_{\mathrm{PL}}$ excited by sawtooth pulses is analogous to $(dV_{\mathrm{ex}})/dt$ [See Fig. S3(e)]. However, in the case of the square pulse [Fig. 2(a)], the response time observed in the waveform of $E_{\mathrm{PL}}$ and $I_{\mathrm{PL}}$ [Figs. 2(f) and 2(g)] is significantly longer than the rise and fall times of $V_{\mathrm{ex}}$, which are $\sim 100$~ps. This may be because the lifetime of a trion is longer than the duration of the electronic disturbance caused by the wave packet (see SM).%\cite{finkelstein}. 

Subsequently, we study the spatial domain. We first focus on the PL spectra at $\Delta t=-2$~ns, in which the wave packet has not yet arrived at the focal point.
The $I_{\mathrm{PL}}$ waveform as a function of $\Delta t$ near the sample edge is overall weaker than that of the QH bulk region [e.g. compare $I_{\mathrm{PL}}$ at $y=0.5$~$\mu$m and $y=1$~$\mu$m in Fig. 3(a)]. 
By moving the focal point towards the QH bulk region along the $y$-axis, $I_{\mathrm{PL}}$ is maximized at $y \sim 1.25$~$\mu$m before decreasing slightly [solid green line in Fig. 3(b)]. $I_{\mathrm{PL}}$ was previously shown to depend on $B$, $\nu$, spin polarization, and $n_e$\cite{hayakawa, moore, yusaPRL, wojs, chklovskii}. 
Owing to the slope of the confining potential [Fig. 3(b) inset], $n_e$ tends to decrease and then became zero beyond the sample edge ($y <0$), contributing to a decrease in $I_{\mathrm{PL}}$ leading to its eventual disappearance. In contrast, the PL peak energy $E_{\mathrm{PL}}$ tends to be constant in two regions, $y< 1$~$\mu$m and $y> 3$~$\mu$m, and monotonically increases between these plateaux by $\sim 0.8$~meV. 
Among other factors, the photon energy of the PL is directly related to the bandgap\cite{hayakawa, moore, yusaPRL, wojs, chklovskii}. In the QH bulk region, the conduction and valence bands in the QW are flat in the $xy$-plane. Thus, $E_{\mathrm{PL}}$ is constant and independent of the spatial position except for small fluctuations owing to the intrinsic random potential on the order of $\sim 100$~$\mu$eV\cite{hayakawa}. 
Closer to the sample edge, the conduction and valence bands tend to shift upward owing to the pinning of the Fermi level by the surface state\cite{chklovskii} (see inset). Therefore, the $y$-component of the electric field $E_y=(dU_{\mathrm{e,h}})/dy$ drifts the electrons and holes toward the QH bulk and sample edge, respectively, and the effective bandgap decreases, resulting in a decrease in $E_{\mathrm{PL}}$. 
The variation in $E_{\mathrm{PL}}$ with $y$ suggests that the slopes of the conduction and valence bands gradually change in the region of $1.5<y<3$~$\mu$m. The fact that $E_{\mathrm{PL}}$ is constant for $y<1$~$\mu$m can be attributed to the almost-constant slopes of the conduction and valence bands.

We consider the PL spectra when a wave packet reaches the focal point. $I_{\mathrm{PL}}$ at $\Delta t=0$ and $1$~ns is weaker and stronger, respectively, than $I_{\mathrm{PL}}$ at $\Delta t=-2$~ns where the wave packet has not yet arrived at the focal point [compare the red and blue solid lines with the green solid line in Fig. 3(b)]. In contrast, $E_{\mathrm{PL}}$ responds to the wave packet differently depending on the $y$ position. $E_{\mathrm{PL}}$ at $\Delta t=1$~ns increases at $y<1.75$~$\mu$m, whereas it decreases at $y>1.75$~$\mu$m [compare red and green dashed lines in Fig. 3(b)].

In Fig. 3(c), we plot the normalized $I_{\mathrm{PL}}$ in the $y$-$\Delta t$ space defined as 
$I_{\mathrm{PL,N}} (\Delta t, y)=\frac{I_{\mathrm{PL}}(\Delta t, y)}
{\overline{I_{\mathrm{PL,0}}(y)}}$, 
where $\overline{I_{\mathrm{PL,0}}(y)}$ denotes the averaged 
$I_{\mathrm{PL}}$ in the range of $-2.3<\Delta t<-1.2$~ns where the wave packet has not yet arrived at the focal point. 
Remarkably, the local minimum in $I_{\mathrm{PL,N}} (\Delta t, y)$  extends several microns into the QH bulk [blue region in Fig. 3(c)] while curving toward a higher $\Delta t$. We observed similar curved features at two other locations relative to the excitation gate ($x=10$ and $25$~$\mu$m, see SM and Fig. S4).
We infer from the finite curvature that the local minimum of $I_{\mathrm{PL,N}}$ penetrating as far as $y>2.5$~$\mu$m in the QH bulk is caused by the influence of the wave packet as it propagates at the edge while also emitting a disturbance directed toward the bulk with an average velocity in the $y$ direction of $v_y \sim 1-3 \times 10^3$~m/s. 
We attribute this disturbance to the magneto-roton collective excitations in the FQH bulk\cite{girvin}. 
The electrons with high energy propagate along the edge, emitting their energy to the bulk.
This can excite the magneto-roton, i.e., the collective excitations of the incompressible bulk in the FQH regime\cite{girvin}.
According to the magneto-roton theory\cite{girvin}, the collective excitation energy of the $\nu =1/3$ Laughlin state is minimized at the wave number $k \sim 1/l_B$\cite{girvin}, where $l_B$ denotes the magnetic length. 
This is called the magneto-roton minimum, and is theoretically expected to be $\sim 0.1 \frac{e^2}{4 \pi \epsilon l_B}$, where $\epsilon$ denotes the dielectric constant of GaAs. 
The velocity of the magneto-rotons $v_{\mathrm{MR}}$ is obtained from their dispersion relation and is $\sim 2 \times 10^4$~m/s, which is an order of magnitude larger than $v_y$ obtained by our experiment. 
This difference can be partly attributed to the finite width of the QW, impurities, and mixing of higher Landau levels, which are known to reduce excitation energies. Because the excitation gap vanishes at the sample edge, further reduction is expected near the edge, wherein the influence of the edge state remains. Therefore, it is reasonable that $v_{\mathrm{MR}}$ near the edge is considerably lower than that predicted in the ideal 2D system.

\begin{figure}[h]
    \par
    \begin{center}
    \includegraphics[clip]{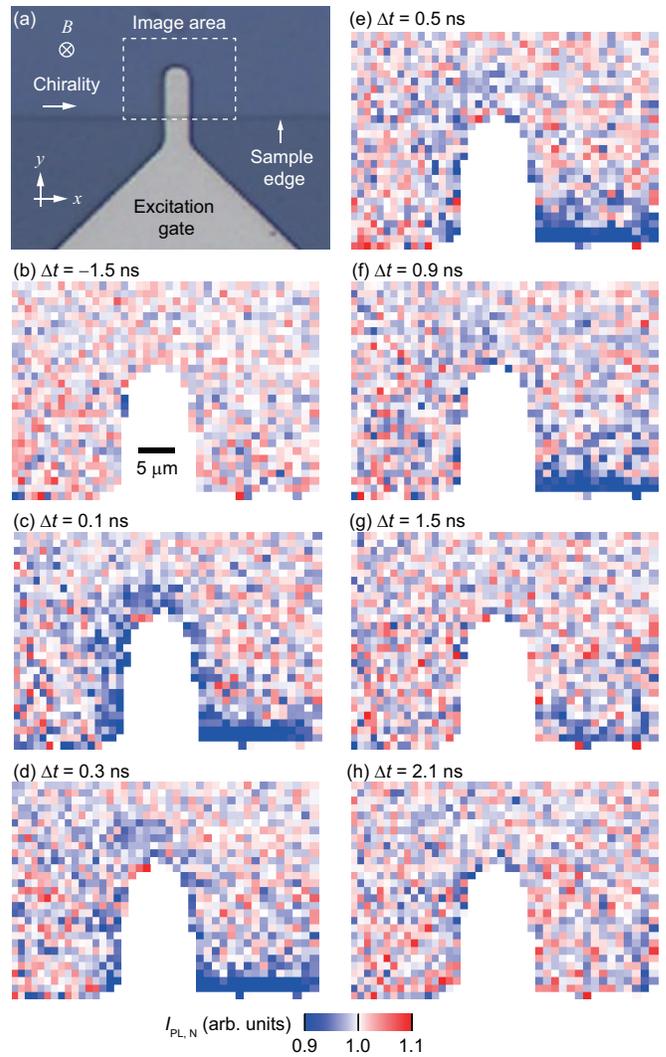}
    \end{center}
    \caption{(Color online) (a) Optical microscope image of the sample near the excitation gate. The $40 \times 29$~$\mu$m$^2$ image area is indicated by the dotted rectangle. (b)-(h) Normalized integrated PL intensity $I_{\mathrm{PL,N}}$ map at $\nu=1/3$ obtained at each $\Delta t$ at $B=11.5$~T and $60$~mK. (See body text for the definition of $I_{\mathrm{PL, N}}$.) $\tau_{\mathrm{pulse}}$ and the amplitude of $V_{\mathrm{ex}}$ are, respectively, $0.5$ ns and $0.3$ V. One pixel is $1 \times 1$ $\mu$m$^2$. See also Movie S2 for the real-time video imaging of $I_{\mathrm{PL,N}}$.}%
    \end{figure}

By raster-scanning the focal point both in the $x$ and $y$-directions and changing $\Delta t$, real-time video imaging of $I_{\mathrm{PL}}$ can be obtained (Movie S1). To focus on the changes in $I_{\mathrm{PL}}$ caused by $V_{\mathrm{ex}}$, we extend the definition of $I_{\mathrm{PL,N}}$ to the $x$-axis, i.e. $I_{\mathrm{PL,N}}(\Delta t,x,y) = \frac{I_{\mathrm{PL}}(\Delta t,x,y)}{\overline{I_{\mathrm{PL,0}}(x,y)}}$. Here, $\overline{I_{\mathrm{PL,0}}(x,y)}$ denotes averaged $I_{\mathrm{PL}}$ in the rage of $-1.5 \leq \Delta t \leq -0.9$ ns where the wave packet has not yet arrived at the focal point $(x,y)$. (See Movie S2 for the real-time video imaging of $I_{\mathrm{PL, N}}$ at $\nu = 1/3$ and Fig. 4 for its temporal slices.) The $I_{\mathrm{PL,N}}$ in the downstream side of the sample edge is clearly decreased at $\Delta t$ from $0.1$ to $0.9$~ns at $\nu = 1/3$ [blue regions in Figs. 4(c)-4(f)]. This suggests that the excitation of the $\nu = 1/3$ edge propagates along the sample edge in accordance with the chirality. 
Notably, a decrease in $I_{\mathrm{PL,N}}$ (blue regions) is also observed around the excitation gate [Fig. 4(c)] and propagates toward the bulk region [Figs. 4(d)-4(f)]. From Figs. 4(c)-4(f), this extends to roughly $\sim 5-10$~$\mu$m for $\sim 0.4-0.8$~ns; thus, the velocity of the wave is roughly estimated to be $\sim 10^{3}-10^{4}$ m/s, which is faster than that observed near the edge ($y<\sim 4$~$\mu$m) [Fig. 3(c)] and agrees with the theoretically expected $v_\mathrm{MR}$, suggesting that magneto-rotons are created by the excitation gate and propagate toward the bulk. 

In summary, we reported on the excited state of the FQH edge and the bulk near the edge in real space and time using polarization-sensitive scanning stroboscopic confocal microscopy and spectroscopy. We observed the behavior of the wave packet propagating along the edge and the bulk region near the edge. The technique demonstrated is a powerful alternative to electrical transport measurements for investigating microscopic dynamical phenomena. It may be adapted to a wide range of other material systems because an optical response, such as reflection spectroscopy, is an alternative for detecting the electronic energy distribution in systems lacking PL. 

\begin{acknowledgments}
The authors are grateful to T. Fujisawa for the fruitful discussions. This work is supported by and a Grant-in-Aid for Scientific Research (Grants Nos. 17H01037, 19H05603, 21F21016, 21J14386, and 21H05188) from the Ministry of Education, Culture, Sports, Science, and Technology (MEXT), Japan.
\end{acknowledgments}

\end{document}